\documentclass[12pt]{article}
\usepackage{epsfig}
\begin{document}

\large
\hfill\vbox{\hbox{DCPT/06/112}
              \hbox{IPPP/06/56}}
\nopagebreak
\vspace{2cm}
\begin{center}
\large{\bf DALITZ ANALYSES: A TOOL FOR PHYSICS WITHIN \\ AND BEYOND THE STANDARD MODEL}
\vspace{8mm}

\large{M.R. PENNINGTON}

\vspace{4mm}
{Institute for Particle Physics Phenomenology, \\Durham University,
Durham DH1 3LE, United Kingdom\\
}
\end{center}
\vspace{6mm}

\small

\begin{abstract}
Dalitz analyses are introduced as the method for studying hadronic decays. 
An accurate description of hadron final states is critical not only to
an understanding of the strong coupling regime of QCD, but also to the 
precision extraction of CKM matrix elements.  The relation of such final state
interactions to scattering processes is discussed.

\end{abstract}

\normalsize
\baselineskip=5.4mm
\parskip=2.mm

\section{Motivation}	

This serves as the introduction to following talks
on Dalitz plot analyses. I will first remind you of the
motivation for such analyses and discuss some aspects of
how to perform them and I leave it to others to describe
detailed results.

We know that there is physics beyond the Standard Model
but we do not yet know what this is. Insight is provided
by precision measurements of the CKM matrix elements,
and in particular the length of the sides and the angles of
the CKM Unitarity triangle. Such measurements in heavy flavour
decays probe
the structure of the weak interaction at distance scales of 0.01fm.
However, the detector centimetres away records mainly pions and kaons
 as the outcome of a typical process like $B \to D (\to K\pi\pi) K$ decay.
Consequently, the very short distance interaction we wish to study is only seen
through a fog of strong interactions, in which the quarks that are
created exchange gluons, bind to form hadrons and then these hadrons interact
for times a 100 times longer than the basic weak interaction.
To uncover this basic interaction we need to understand the 
nature of these strong processes, or at least to model them
very precisely. Indeed, the biggest uncertainties in determining the
CKM angle $\phi_3$, or $\gamma$, from the difference of $D$ and ${\overline D}$
decays is due to our inability to model the final state interactions~\cite{cleoc1,belle1,babar1}.
Dalitz plot analyses are the way to improve this.

What helps here is that the most common particles for $B$'s, $D$'s,
and even $J/\psi$'s and $\phi$'s, to decay into are pions and kaons.
These, being the lightest of all hadrons, have final state interactions
that are common to all these processes. So each can teach us about the other
and in turn heavy flavour decays are now the richest source of
information about light quark dynamics.
\vspace{-3mm}
\begin{figure}[h]
\centerline{\psfig{file=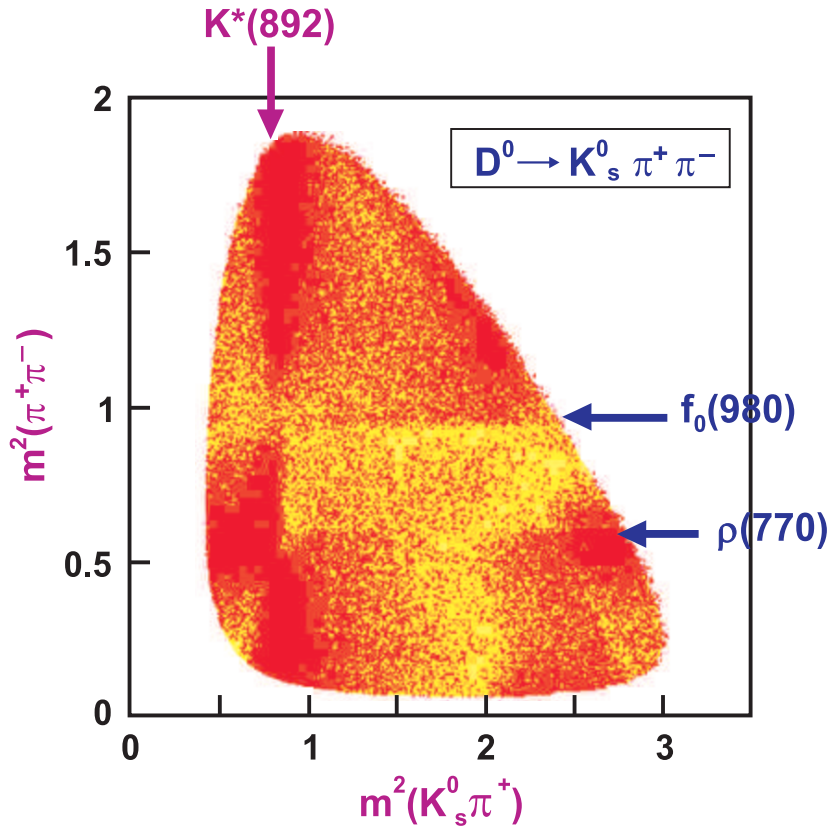,height=10.9cm,width=10.9cm}}
 \caption{\small{Example of a Dalitz plot, here for $D^0\to K_s\pi^+\pi^-$ decay.}}
\end{figure}

As soon as one has more than two particles in the final state,
the events are most readily pictured in a Dalitz plot, an example of which 
is shown in Fig.~1. This was invented by
Richard Dalitz (known to everyone as {\it Dick}) more than 45 years ago~\cite{dalitz}.
Sadly Dick died earlier this year aged 81 and this talk is dedicated to him.
  For a three body decay, like $D \to K\pi\pi$, that we will concentrate
on here, the data are plotted with the mass squared 
of the $K\pi$ and $\pi\pi$ systems on the $x,y$ axes, Fig.~1. Now the first thing to notice about
such a plot is that the events are not uniformly distributed.
The decay does not proceed by $D$ decaying to the three body system $K\pi\pi$
directly, rather there are a series of bands which show that
the $D$ likes to emit a $K$ and then form a resonance like the $\rho$,
which later decays into two pions, or the $D$ emits a $\pi$ forming
a $K^*(892)$, which then later decays to $K\pi$. Thus to describe the
basic matrix element for the decay we need to know how to represent
the vector mesons like $\rho$ and $K^*$. These, being reasonably long lived,
have a magnitude and phase change across the resonance well described
by a simple Breit-Wigner formula, with a pole in the complex energy plane on
the nearby unphysical sheet. However, this represents only a small
fraction of the events in the Dalitz plot.

The simplest way for any state of heavy flavour to lose mass
is to emit a $\pi$ or $K$ and form a scalar meson. Having $J^{PC}=0^{++}$
 this produces no change in angular momentum and so is almost cost free.
Indeed, any system can lower its mass by forming a scalar with $I=0$, 
since this has vacuum quantum numbers.
 This we can study in $\pi\pi$ scattering.
 There we see that the cross-section, Fig.~2, does not appear
 to have any structures
looking like simple Breit-Wigners, rather it has a series of broad peaks with
deep narrow dips between~\cite{zou}.
\vspace{5mm}
\begin{figure}[h]
\centerline{\psfig{file=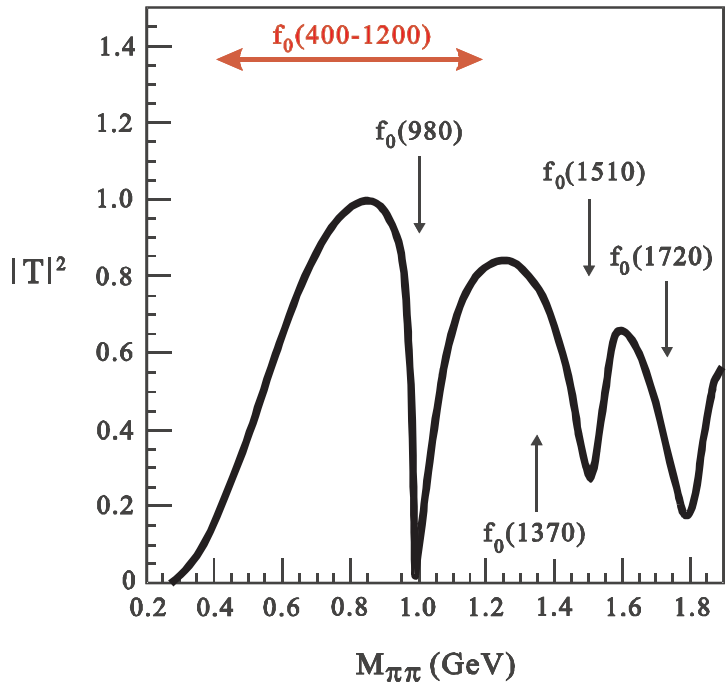,width=9.5cm}}
\vspace*{8pt}
 \caption{\small{Sketch of the modulus squared of the $I=J=0$ $\pi\pi\to\pi\pi$ amplitude from Ref.~5. 
Resonances in the PDG tables~\protect\cite{pdg} are indicated.}}
\end{figure}
 The increase from threshold is related 
to the short-lived $\sigma$ and the dramatic dip at ${\overline K}K$ threshold
 to the $f_0(980)$.
Neither can be represented by a simple Breit-Wigner and certainly naively summing 
Breit-Wigners would violate the conservation of probability.
Consequently, we will need a better way to treat these broad and overlapping 
structures throughout the $D$ and $D_s$ decay regions.
Indeed data on these decays provide unique information about the
strong coupling regime of QCD, which defines the
Higgs sector of chiral symmetry breaking. Even if we do not care about
this key aspect of the Standard Model,
we nevertheless need an accurate way to parametrize its effects to determine
the CKM matrix elements.

\begin{figure}[h]
\centerline{\psfig{file=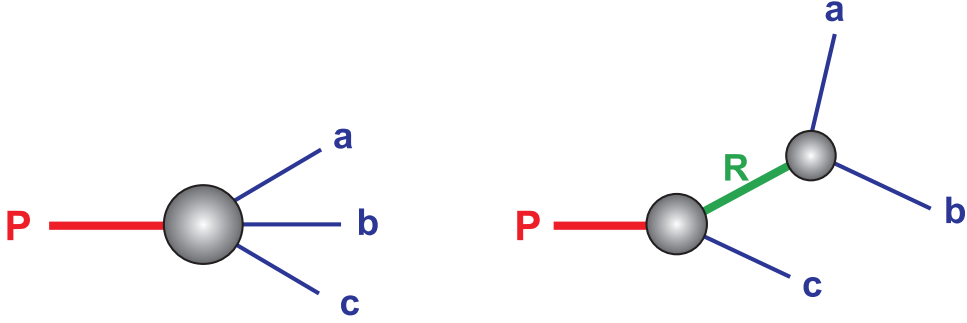,width=12.8cm}}
\vspace*{8pt}
 \caption{\small{ Decay of a parent particle $P$ to $abc$. $R$ is a resonance in the $ab$ channel, which is the basis of the isobar model of sequential decays: $P\ \to Rc$, then $R \to ab$, plus contributions $P\ \to R' a$, then $R'\to bc$ and $P\ \to R'' b$ and $R'' \to ca$ with $c,\ a$ and $b$ as spectators, respectively.}}
\end{figure}

We have seen that much of 3-body decays proceeds as a two stage process.
This is embodied in the isobar picture, Fig.~3. There one assumes 
that parent particle $P$ spits out a particle $c$, for instance, to produce a resonance
$R$ that subsequently decays into $ab$, particle 
$c$ being regarded as a spectator
as far as the final state is concerned. Such a picture immediately
implies a modelling for the process $ab \to ab$, in which the same resonance
$R$ appears. One must check that these descriptions are consistent.
The resonance must have the same mass and total width, as well
as partial width to the $ab$ channel. This constraint is encoded in
unitarity, which enforces the conservation of probability.
Indeed, unitarity does not differentiate between the resonance and 
its background, but treats them both together. This is particularly
important for short-lived states like the $\sigma$, where such a distinction
is totally semantic. Thus unitarity requires that if the particle $c$ is
a spectator then for instance the imaginary part of the amplitude
for $P \to \pi\pi(c)$ is equal to the sum shown in Fig.~4, where one sums
over all kinematically allowed hadronic intermediate states. 
This means that if we
represent the scattering amplitude for these intermediate processes
$ab \to h_n$ by the matrix $T$, then the amplitude for the decay
$F$ is given by the vector equation
\begin{eqnarray}
F(P\to ab(c))&=&\sum_n\,\alpha_n\,T(h_n\to ab)\quad ,
\end{eqnarray}
where the functions $\alpha_n$ represent the basic coupling
of $P\to h_n (c)$ for each hadronic intermediate state $h_n$.
\begin{figure}[h]
\centerline{\psfig{file=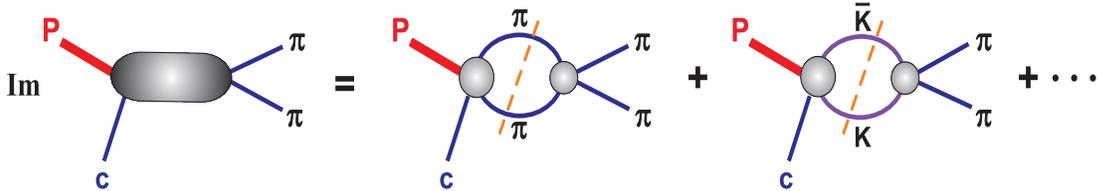,width=14.5cm,height=2.5cm}}
\vspace*{4pt}
 \caption{\small{ Unitarity for the $ab$ system in the decay $P\to (ab)c$, where $c$ is a spectator. The dashed lines denote the particles in the intermediate state are on mass shell.}}
\end{figure}

 When $c$ is a spectator these coupling functions
are real. The hadronic amplitudes $T$ can be conveniently
represented by a $K$-matrix form to ensure unitarity for the
hadronic amplitudes (or by related forms I call the $L$-matrix
if we are also to require the correct analyticity).
At the lowest energies when $ab$ is the only accessible channel
this implies Watson's theorem~\cite{watson}, namely that the phase of
each decay amplitude (with definite isospin and spin),
must equal that for the corresponding hadronic scattering amplitude for $ab \to ab$.

\section{Semi-leptonic decays}

Now we can test this by considering the semileptonic decay
$D^+\to K^-\pi^+ (\mu^+\nu_{\mu})$, where clearly only
the $K\pi$ system can have strong
final state interactions.
The dominant low energy signal is in the region of the $K^*(892)$,
so FOCUS~\cite{focusdl4} looked at the mass region from 800 MeV to 1 GeV in two bins.
If the vector $K^*$ was all there was then the 
angular distribution of the final state hadrons in the $K\pi$ rest system 
would be proportional to $\cos^2 \theta$ and so forward-backward symmetric.
However a marked asymmetry is found, Fig.~5. This means the $P$-wave $K\pi$
amplitude must interfere with some other wave and at low energies that is
the $S$-wave. It is here that many have argued lies the scalar $\kappa$.
What does experiment tell us?
\begin{figure}[h]
\vspace{-3mm}
\centerline{\psfig{file=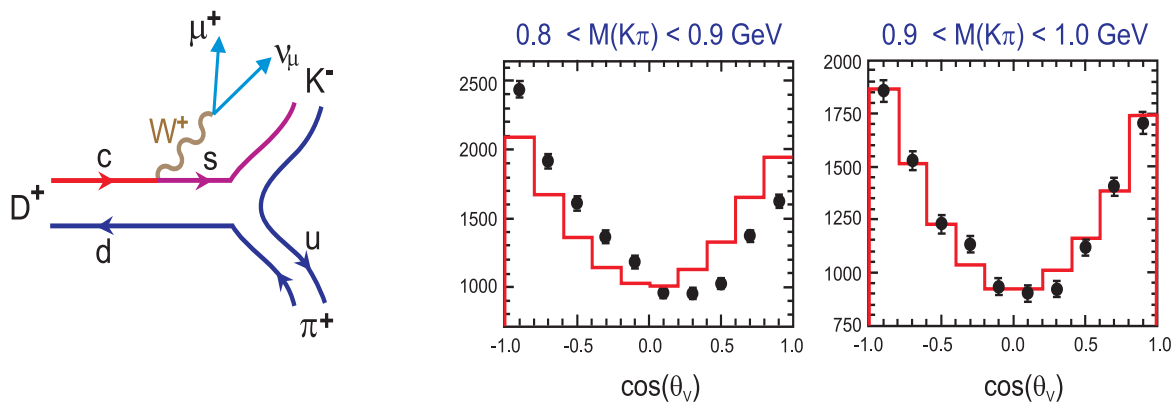,width=15.5cm,height=5.5cm}}
\vspace*{6pt}
 \caption{\small{ Feynman graph of the semi-leptonic decay of a $D^+$. Asymmetry in $K\pi$ angular distribution in the $K^*(892)$ mass region from the FOCUS experiment~\protect\cite{focusdl4}.}}
\end{figure}

Let us first look at $K\pi$ elastic scattering. This we learn about
by studying high energy $K\pi$ production with $K$-beams at small momentum 
transfers, where the reaction is controlled by one pion exchange. 
Then one can extract the
cross-section for $K\pi \to K\pi$ scattering. This was done 25 years
ago by LASS~\cite{lass}. The cross-section shows clear $K^*(892)$ and $K_2^*(1430)$
peaks. By performing a partial wave analysis~\cite{lass},
 one confirms that these
have spin-1 and 2 respectively with the magnitude and phase of well-defined
resonances. What we are interested in here is the $S$-wave underneath these.
This shows a broad rise in magnitude and phase from 825 MeV (where data
begin) with a peak at 1400 MeV and phase rising through 90$^o$. This
is characterised by the $K_0^*(1430)$ with 
a width of $\sim 300$ MeV. There is no hint
 of a low mass $\kappa$, which if it were simply 
describable by a Breit-Wigner, 
would have forced the phase to have already reached 90$^o$ by 800 MeV.
What does $D$ semileptonic decay tell us? FOCUS~\cite{focusdl4}
 found that their
forward-backward asymmetry, Fig.~5,
 requires an $S$-wave phase of 45$^o$ in the 
$K^*(892)$ region, exactly as LASS has measured~\cite{lass}. 

The statistics at FOCUS only allow a determination 
of the $K\pi$ $S$-wave phase
in the $K^*(892)$ region and then only in  very wide bins.
However, such $D$ semileptonic decays with the event rates of CLEO-c and 
$B$-factories should provide an accurate determination of the low energy
$K\pi$ phase-shifts, in the same way as $K_{e4}$ decays have done for
$\pi\pi$ phases. These, when combined with dispersion relations and  three channel crossing symmetry,
 have resulted in a rather
precise determination of the position of the $\sigma$-pole~\cite{ccl}.
If we are going to be able to determine whether and where a $\kappa$
exists, then we will need precision $K\pi$ information below the range accessed
by LASS. Semileptonic $D$ decays are the theoretically unambiguous way to go.

\section{Hadronic final states}
While we wait for that, let us turn to 3-hadron decays and let us focus on
the Cabibbo-favoured $D^+\to K^-\pi^+\pi^+$ channel.
Let us consider this with increasing
 levels of sophistication. First let us assume
an isobar picture. The Dalitz plot is to be described
by a sum of isobars in the three di-meson channels, Fig.~3. Since the $\pi\pi$
channel has $I=2$, it has no known resonances and so is set to zero.
The $K\pi$ channels are described simply by summing Breit-Wigners with
parameters taken from the PDG tables~\cite{pdg}, the $K^*(892)$, $K_1^*(1410)$,
$K_2^*(1430)$ and of course the broad $K^*_0(1430)$. To this is added
the 3-body interaction matrix element, which is presumed constant in both
magnitude and phase across the Dalitz plot. The resultant fit to the E791
data~\cite{e791brian} is very poor and 90\% of the decay is ascribed to the direct 3-body
term. Since the Dalitz plot displays distinct 2-body structures, no
wonder the fit is poor.
The next step is to add a Breit-Wigner for another scalar, 
they call the $\kappa$. It is just added. Its mass and width 
are determined by the fit~\cite{e791kappa} to be $M=797$ MeV, $\Gamma=410$ MeV. The fit dramatically 
improves. The direct 3-body fraction is down to a more believable 14\%, 
and the $\kappa$ contributes 43\%. However modelling the scalar channel
by a sum of the $\kappa$ and $K^*_0(1430)$ Breit-Wigners is not only 
in violation of the conservation of probability, but implies a model for
$K\pi$ elastic scattering in total disagreement with the LASS results~\cite{lass}.
We clearly have to do better.

Rather than be tied to specific forms for the complicated $S$-wave amplitude,
Brian Meadows~\cite{e791brian,e791pub} has tried something much more promising.
With $P$ and $D$-wave $K\pi$ interactions given as before by sums of 
Breit-Wigners from the PDG tables~\cite{pdg}, the $S$-wave is parametrized in terms of a magnitude 
and a phase in each bin across the Dalitz plot. Fitting gives the phases 
 shown in Fig.~6.
These are determined almost to threshold and, if they had greater precision and
we applied the right tools for analytic continuation, might result in locating a
$\kappa$ pole on the nearby unphysical sheet. So how do these phases
compare with the $K\pi$ elastic $S$-wave? 

\begin{figure}[h]
\centerline{\psfig{file=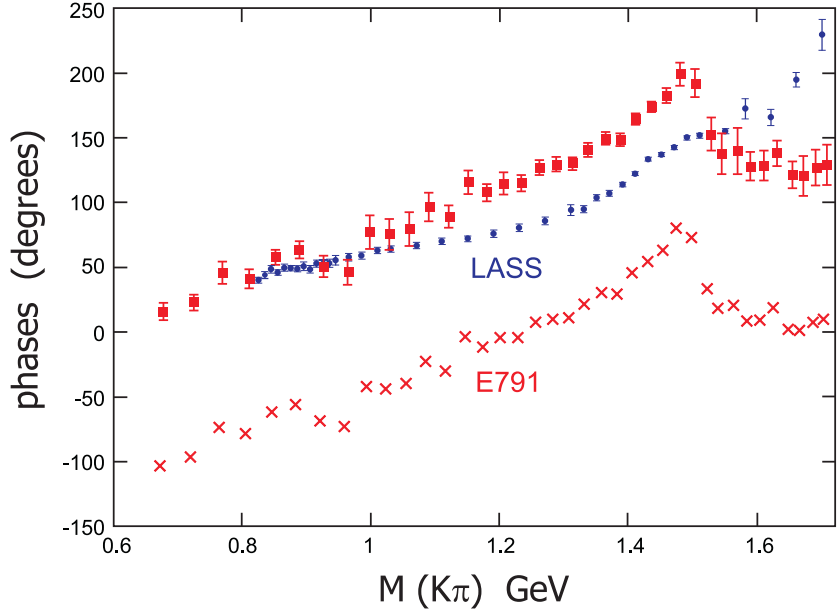,width=11.cm}}
\vspace*{6pt}
 \caption{\small{ The points labelled LASS are the $S$-wave phase for $K^-\pi^+\to K^-\pi^+$ from analysis of the
 LASS experiment~\protect\cite{lass}. The solid data points are the $K^-\pi^+$ $S$-wave phase from Meadows' 
analysis~\protect\cite{e791brian,e791pub} of
the E791 results on $D^+\to K^-\pi^+\pi^+$ decay shifted from their original position, marked by crosses, by $\sim 100^o$. 
}}
\end{figure}

In Fig.~6 the $K\pi$
phases from Brian Meadows' E791 fit~\cite{e791pub} and the phases 
for $K^-\pi^+\to K^-\pi^+$
scattering from LASS~\cite{lass} are shown. 
The latter are absolute. Those from $D$-decay are 
relative, relative to the $P$-wave fixed to be 90$^o$ at 892 MeV. 
Consequently, we are free to raise the E791 phase up to be zero at threshold.
 We then see better agreement, with a common rising trend with $K\pi$ mass.
However, Fig.~6 shows these are clearly not the same. There are several possible reasons
 for this. $K^-\pi^+$ interactions involve both $I=1/2,\,3/2$ components.
In elastic scattering the relative strength of these is fixed by Clebsch-Gordan
coefficients, while in $D$-decay this is determined by dynamics. We may have the
prejudice that $I=1/2$ is dominant, but that does not mean  the 
$I=3/2$ component is negligible. Thus if we assume Watson's theorem, which holds in the elastic region effectively up to $K\eta'$ threshold, one can 
determine the relative amounts of $I=1/2$ and $I=3/2$
contributions, as in Ref.~14. However, what I want to present here is 
something a little different.

\begin{figure}[h]
\centerline{\psfig{file=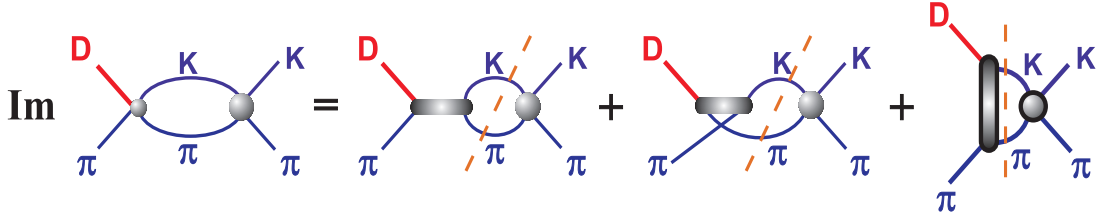,width=14.5cm,height=2.8cm}}
\vspace*{2pt}
 \caption{\small{ Unitarity for the $K\pi$ system in $D$-decay in the elastic region.
The dashed lines denote the particles in the intermediate state are on mass shell.}}
\end{figure}

So far we have considered the final state interaction of the $K$ and a $\pi$
where the second pion is merely a spectator. Let us now ask what happens
if we try to include a subsequent interaction of the $K$ with this 
pion. There has for long been a body of work on such multiparticle interactions~\cite{aitchison}, particularly
by Ascoli and collaborators~\cite{ascoli} on 3 pion final states dating from the 
discovery of the $a_1$ and its possible structures. Such multiparticle final states have been investigated by Anisovich {\it et al.}~\cite{anisovich} in $\overline{p}p$ annihilations at LEAR. Much more recently Caprini~\cite{caprini}
 has shown 
that one can deduce a unitarity relation with rescattering, implicit in these studies.
This relation shows that for each partial wave amplitude 
the imaginary part of the $D\to K\pi\pi$ is a sum of 
contributions, Fig.~7.

\begin{figure}[h]
\centerline{\psfig{file=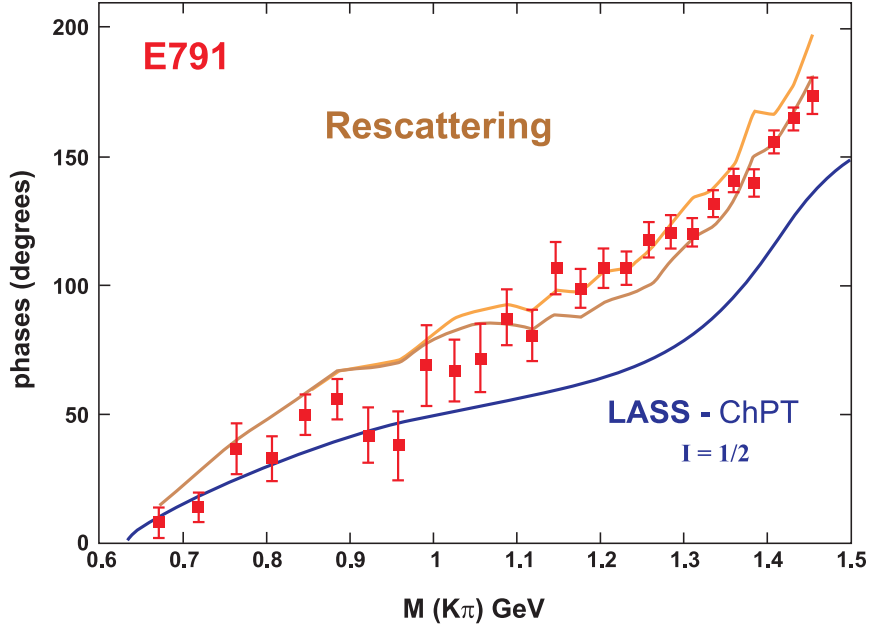,width=11.4cm}}
\vspace*{5pt}
 \caption{\small{ The solid curve denotes a fit to the LASS results in Fig.~6 consistent with Chiral Perturbation Theory~\protect\cite{descotes} for the $I=1/2$ $S$-wave $K\pi$ phase compared
with the $S$-wave $K\pi$ phase from the Meadows' analysis~\protect\cite{e791brian,e791pub} of the E791 results
on $D\to K\pi\pi$ decay. The upper line displays a preliminary calculation 
of how the elastic scattering phase incorporated in the first graph on the right hand side of the unitarity relation shown in Fig.~7 is modified by the rescattering corrections
from the second (and third) graphs in Fig.~7. The curve closest to the E791 phases includes feedback.}}
\end{figure}

If there was just the first term in Fig.~7 then Watson's theorem would hold in the region of elastic unitarity. The second and third graphs in Fig.~7 give corrections.
Since the Dalitz plot for $D^+\to K^-\pi^+\pi^+$ is symmetric in the two $K\pi$ systems, and assuming the $\pi^+\pi^+$ amplitude to be negigible, we can then use the Meadows' results to compute the rescattering corrections given by these graphs.
There are some technicalities I won't go into here, but the crude result
gives the phase of the $S$-wave $K^-\pi^+$ interaction in $D$-decay to be the upper line in Fig.~8. The corrected phase is computed at each datum  of the Meadows' analysis and then these are simply connected by straight lines. The similar line a little bit lower (and closer to the data) involves including feedback to improve the
phase variation across the Dalitz plot.

These preliminary results are encouraging. However, details still need to be checked before they can be considered definitive. Nevertheless, this method holds out the prospect of using high statistics data on hadronic $D$-decays from BaBar and Belle to determine the near threshold $K\pi$ phase-shift with precision, independently of summing higher orders in Chiral Perturbation Theory. So while there is no $\kappa(900)$~\cite{cherry}, there may be a scalar much closer to threshold deep in the complex plane. From the Roy equation analysis of $\pi\pi$ scattering we now know the position of the $\sigma$-pole rather precisely. All analyses should find the same result. The variation between treatments discussed in Refs.~20, 21 is unacceptable in the era of precision physics. The application of the rescattering corrections to $J/\psi\to\omega\pi\pi$ may well, with higher statistics accessible at BESIII, bring the mass and width from a simple Breit-Wigner treatment~\cite{bes2} in line with the true $\sigma$-pole position~\cite{ccl}. That is the challenge for BES.

A precise description of these hadronic final state interactions, 
especially those in scalar channels, is essential to reducing the uncertainty 
in the CKM triangle. This reduction is crucial to learning
about physics both within and beyond the Standard Model.  Dalitz analyses are at the heart of this programme, as others will describe.

\section*{Acknowledgments}
It is a pleasure to thank the organisers and members of IHEP,
particularly Yifang Wang, Shan Jin, Haibo Li and Medina Ablikim,
 for their kind hospitality.
I am grateful to the UK Royal Society for a conference travel grant. This work
was supported in part by the EU-RTN  Programme,
Contract No. HPRN-CT-2002-00311, \lq\lq EURIDICE''.

\end{document}